\newcommand{\ax} {\`{a} }

\newcommand{\bee} {\begin{equation}}
\newcommand{\ene} {\end{equation}}
\newcommand{\beqa} {\begin{eqnarray}}
\newcommand{\enqa} {\end{eqnarray}}
\newcommand{\beqsa} {\begin{eqnarray*}}
\newcommand{\enqsa} {\end{eqnarray*}}
\newcommand{\bea} {\begin{array}}
\newcommand{\ena} {\end{array}}

\documentstyle[12pt]{article}
\begin{document}
\baselineskip=18pt

\begin{center}
{\Large\bf Quantum statistics and Altarelli-Parisi evolution equations}\\
\end{center}

\bigskip\bigskip

\begin{center}
{\bf G. Mangano}, {\bf G. Miele} and {\bf G. Migliore}
\end{center}

\bigskip

\small
\noindent
{\it Dipartimento di Scienze Fisiche, Universit\ax di Napoli - Federico II -,
and INFN - Sezione di Napoli, Mostra D'Oltremare Pad. 20, 80125, Napoli,
Italy.}

\normalsize

\bigskip\bigskip\bigskip

\begin{abstract}
The phenomenological evidence of quantum statistical effects in parton physics
is here briefly summarized, and the recent good results obtained by
parameterizing the parton distributions in terms of Fermi-Dirac and
Bose-Einstein statistical functions are discussed. In this framework we study
the modification of the scaling behaviour of parton distributions due to
quantum statistical effects. In particular, by following a well-known formal
analogy which holds between the Altarelli-Parisi evolution equations, at
leading-log approximation, and a set of Boltzmann equations, we suggest a
generalization of evolution equations to take into account Pauli exclusion
principle and gluon induced emission.
\end{abstract}

\vspace{2cm}
\begin{center}
{PACS number: 13.60.-r}
\end{center}

\vspace{2cm}

\centerline{\it published in Nuovo Cim. A108 (1995) 867-882.}
\newpage

\section{Introduction}

The low $x$ regime in deep inelastic processes has recently received much
attention due, in particular, to the advent of the HERA electron-proton
machine, which would provide precision measurements in the region $Q^2 >
10~GeV^2$ and $x \geq 10^{-4}$ \cite{wolf}. In the limit of very small momentum
fraction one deals with a dense system of partons in a weak coupling limit, in
which, however, interactions among partons cannot be neglected, being able to
build up, as we know from many cases in condensed matter physics, a collective
dynamics. The aim of this paper is to study the effect of statistical
correlations among partons, due to their Fermi or Bose nature, in the $Q^2$
evolution of their distribution functions. These correlations, in fact, would
be expected whenever the parton wave functions overlap.

In terms of the two {\it phase-space} variables $Q^2$ and $x$, it is possible
to distinguish three regions in which strong interactions among partons,
dictated by QCD, behave quite differently:

\bigskip

\noindent
1) For high values of $Q^2$ and small densities $\rho$, defined as the
number $N$ of partons per unit of rapidity $y=\log(1/x)$ in the
transverse plane
\begin{equation}
\rho = { d N \over dy} ~{ 1 \over \pi R_{h}^2}
\label{01}
\end{equation}
with $R_{h}$ the radius of the hadron, one can powerfully apply the
perturbative QCD methods. In particular, the $Q^2$ evolution of
structure functions can be evaluated at leading-log level in $Q^2$ by
standard Altarelli-Parisi equations (AP) \cite{ap} when the following
conditions are satisfied:  $\alpha_{s} <<1$,
$\alpha_{s}~\log{Q^2} \approx 1$, and
$\alpha_{s}~\log(1/x) <<1$. Alternatively, in the kinematical region:
$\alpha_{s}~\log{Q^2} <<1$, and $\alpha_{s}~\log(1/x) \approx 1$ and
still with $ \alpha_{s} <<1$, the
approach of extracting the contribution of the order
$\left[\alpha_{s}~\log(1/x)\right]^n$, leads to Fadin-Kuraev-Lipatov
equations (FKL) \cite{fkl}.

\bigskip

\noindent
2) The low $Q^2$ regime, or equivalently the long-distance interaction
region, is typically the realm of non-perturbative QCD. In this case,
the value of the strong coupling constant is large and one is
dealing with the confinement problem.

\bigskip

\noindent
3) Finally,
for high $Q^2$ and large densities, as already mentioned, we are still in
non-perturbative conditions, but in this case, the latter are rather due to the
large number of partons which interact each other. This high density QCD regime
is particularly interesting from a theoretical point of view, since the small
value of the coupling constant gives us a chance to successfully face the
problem. Many attempts have been already addressed to understand the main
characteristics of this parton-plasma dynamics. In particular, due to
interactions among partons, one should expect nonlinear effects in the $Q^2$
evolution of distribution functions: stated differently, microscopical
processes with two or more partons in the initial state become relevant in
changing their resulting number. This point of view is at the basis of
Gribov-Levin-Ryskin equations (GLR) \cite{glr}, where besides parton decays,
whose probability is proportional to $\alpha_{s} \rho$, parton annihilation
processes are explicitly taken into account. These introduce in the scaling
equations a quadratic term in the distribution functions of the form
$\alpha_{s}^2 \rho^2/Q^2$. Hence, one gets for the particle balance in a cell
of the phase-space a {\it Vlasov} equation
\begin{equation}
{ \partial^2 \rho \over \partial y \partial \log{Q^2}} =
{3 \alpha_{s} \over \pi} \rho - {\alpha_{s}^2 \over Q^2 } \rho^2 \cdot
{}~{\mbox{const.}}~~~.
\label{02}
\end{equation}

One comment is in order at this point: the nonlinear effects expected in the
evolution can be both of {\it dynamical} and/or {\it statistical} nature. The
first ones  are simply due to interaction processes among partons which are
very close each other, and are the ones included to some extent in Eq.
(\ref{02}). However, in a dense medium quantum statistics may provide similar
relevant effects, related in particular to exchange interactions for Fermi
particles (quarks) and induced emission processes for Bosons (gluons). Despite
of this in the literature this point has not been sufficiently remarked. The
decay process, for example, of a parton into a quark with a definite momentum,
if it occurs in presence of many other quarks with the same momentum, would be
strongly suppressed. In this case, the corresponding probability would be not
only proportional to the decaying parton distribution function, but also to a
{\it Pauli blocking} factor, which depends on the final quark density.

In order to find a simple way in which the statistical effects can be taken
into account in the evolution equations, we will start from a phenomenological
description of parton distributions in terms of {\it equilibrium-like}
functions \cite{bour}. This {\it thermodynamical} approach to deep inelastic
scattering phenomena was developed on the basis of previous papers \cite{buc},
where Gottfried sum rule violation \cite{got}, \cite{nmc} and other typical
behaviour of structure functions are interpreted in terms of Pauli exclusion
principle. In section 2 we will briefly review this idea, showing the good
agreement of the theoretical predictions for unpolarized and polarized
structure functions with NMC, EMC and E142 experimental data
\cite{nmc}, \cite{emc}, \cite{e142}. This agreement is not completely
surprising if one
stresses the fact that, as shown in \cite{coll}, the $Q^2$-evolution can be
consistently viewed as a {\it thermalization} process. According to this
analysis, the AP scaling equations at leading-log can be shown to be equivalent
to a set of Boltzmann transport equations, which, as well-known, describe the
approach towards equilibrium conditions
of a thermodynamical system, where a simple
function of $Q^2$ plays the role of time. In this sense one can guess, in
strict analogy with H-Boltzmann theorem, that parton distributions would
asymptotically reach equilibrium shape at infinite $Q^2$. The good agreement
with the data of equilibrium-like distributions at $Q^2=4~GeV^2$, could
therefore suggest that thermalization process is quite rapid.

The stated analogy between transport equations and AP straightforwardly leads
to a generalization of the latter in a regime of quite large densities
\cite{Fermilab}. In fact, AP are strictly equivalent only to a set of Boltzmann
equations for a very dilute system, where all quantum statistical effects,
namely Pauli blocking and induced gluon emission, are negligible. These can be
simply introduced by adding in the {\it collisional integral} appropriate
factors of the form ($1 \pm f$), where $f$ are parton statistical functions,
namely their distributions once the {\it level degeneracy} has been
subtracted out (see section 3). This procedure leads to a set of generalized
nonlinear AP equations, which recover the usual AP in the low
density region, but whose validity is quite wider, since the effects originated
by quantum statistics have been explicitly taken into account.

The paper is organized as follow: in section 2 we briefly review main
experimental results leading to the conclusion that quantum statistics may play
a role in parton dynamics inside hadrons. The {\it thermodynamical} model
proposed in \cite{bour} is also described. In section 3 we will show, following
\cite{coll}, that AP equations can be formally viewed as transport equations:
this can be achieved by considering non regularized splitting functions and by
explicitly computing  infrared virtual gluon contributions. The generalization
of AP to a new set of nonlinear equations which contain quantum statistics
effects is the subject of section 4. Finally in section 5 we give our
conclusions and remarks.

\section{Pauli exclusion principle in deep inelastic scattering}

\subsection{Experimental results}

Deep inelastic experiments seem to be an inexhaustible source of information on
the hadronic structure and continue to considerably improve our understanding
of strong interaction dynamics. A measurement of proton and neutron $F_{2}(x)$
structure function performed by the NMC Collaboration at CERN \cite{nmc}
suggests a rather large $SU(2)$ flavour breaking in the sea quark \cite{prep}.
In particular they have obtained a determination for the difference
\bee
{\cal I}_{G} = \int_{0}^{1} \frac {dy}{y} [ F_{2}^{p}(y)-
F_{2}^{n}(y)] = 0.235 \pm 0.026~~~,
\label{eq:exgsr}
\ene
instead of the value $1/3$ predicted by an $SU(2)$ symmetric sea; in fact
\bee
{\cal I}_{G} = \frac{1}{3}~(u+{\bar u}-d-{\bar d})=
\frac{1}{3} + \frac{2}{3}~ (\bar{u} - \bar{d})~~~.
\label{eq:gsr}
\ene
This result, which represents a relevant violation of the Gottfried sum rule
\cite{got}, yields
\bee
\bar{d}-\bar{u} = \int_{0}^{1} dx~ [ \bar{d}(x)- \bar{u}(x) ] = 0.15
\pm .04~~~.
\label{eq:du}
\ene
The inequality $\bar{d}>\bar{u}$, however,  was already argued many years ago
by Field and Feynman \cite{ff} on pure statistical basis. They suggested that
in the proton the production from gluon decays of $u \bar{u}$-pairs with
respect to $d \bar{d}$-pairs would be suppressed by Pauli principle because of
the presence of two valence $u$ quarks but of only one valence $d$ quark.
Assuming this point of view, the experimental result (\ref{eq:exgsr}) naturally
leads to the conclusion that quantum statistical effects would play a sensible
role in parton dynamics and that, in particular, parton distribution functions
are affected by them. In this picture one may also easily account for the known
dominance at high $x$ of $u$-quarks over $d$-quarks, whose characteristic
signature is the fast decreasing of the ratio $F_{2}^{n}(x)/F_{2}^{p}(x)$ in
this regime. Fermi statistics imply, in fact, a broader distribution for $u$
quarks, due to their larger abundance.

Another evidence for the effect of the Pauli principle on the parton structure
follows from the double helicity asymmetry for polarized muon (electron) -
polarized proton deep inelastic scattering $A_{1}^{p}(x)$.
By denoting with $q^{+}(x)$
($q^{-}(x)$) quark distributions with helicity parallel (antiparallel) to the
proton helicity, $A_{1}^{p}(x)$ is defined as
\begin{equation}
A_{1}^{p}(x) \equiv { g_{1}^{p}(x) \over F_{1}^{p}(x) } \approx
{ 4 [ u^{+}(x) - u^{-}(x) ] + [ d^{+}(x) - d^{-}(x) ] \over
 4 ~u(x) + d(x)}~~~.
\label{eq:a1p}
\end{equation}
Experimentally this quantity increases towards unity
for high $x$ \cite{emc}, thus in this
regime $u^{+}(x)$ dominates over $u^{-}(x)$, $d^{+}(x)$, and $d^{-}(x)$. This
interesting behaviour can be interpreted reminding that at $Q^{2}=0$ the first
momenta of the {\it valence} quark distributions are related to the axial
couplings $F$ and $D$ through the following relations
\bee
u^{+}_{{val}}=1+F~,~~~~~u^{-}_{{val}}=1-F~,~~~~~
d^{+}_{{val}}={ 1 + F - D \over 2}~,~~~~~
d^{-}_{{val}}={ 1 - F + D \over 2}~~~.
\label{eq:val}
\ene
Reminding that $F=0.477 \pm .011 \approx 1/2$ and $D=.755 \pm .011 \approx 3/4$
\cite{bourquin}, we get for the valence quark abundances $u^{+}_{{val}} \approx
3/2$, $u^{-}_{{val}}\approx 1/2$, $d^{+}_{{val}} \approx 3/8$ and
$d^{-}_{{val}} \approx 5/8$. The fact that dominant distributions correspond to
highest values of the valence abundances gives the {\it abundance - shape}
correlation, which is the typical property of the Fermi - Dirac distribution
function: larger abundances correspond to broader distributions.  In
particular, from the previously obtained values for the first momenta one can
extrapolate the useful relation valid for the quark distributions \cite{buc}
\bee
u^{-}(x) = { 1 \over 2} d(x)~~~,
\label{eq:dq}
\ene
which leads to
\bee
\Delta u(x) \equiv u^{+}(x) - u^{-}(x) = u(x) - d(x)~~~.
\label{eq:duq}
\ene
This equation allows to relate the contribution in the proton polarized
structure function $g_{1}^{p}(x)$ due to the $u$ quarks to the one due to $u$
and $d$ present in $F^{p}_{2}(x) - F^{n}_{2}(x)$, i.e.
\bee
x g_{1}^{p}(x)\Bigr|_{u} ~\approx~ \left.
{ 2 \over 3}\left(F^{p}_{2}(x) -
F^{n}_{2}(x) \right)\right|_{u+d}~~~.
\label{eq:xgp-f2pn}
\ene
Then, neglecting the $d$ quarks term in $g_{1}^{p}(x)$ ($\Delta d_{{val}} =
-1/4~\Delta u_{{val}}$ and $e_{d}^2 = 1/4~e_{u}^2$), we get
\bee
x g_{1}^{p}(x)~\approx~ { 2 \over 3}\left(F^{p}_{2}(x) -
F^{n}_{2}(x) \right)~~~,
\label{eq:xgp-f2pn1}
\ene
at least in the region dominated by valence quarks. This relation is in
good agreement with the experiment \cite{nmc}, \cite{emc}.

\subsection{Quantum statistical approach to parton distributions}

In a recent paper \cite{bour}, the idea to extensively consider Pauli
principle and use Fermi-Dirac and Bose-Einstein statistics for the parton
distributions has been developed. It has succeeded in making reasonable
assumptions for various polarized parton distributions in terms of unpolarized
ones, explaining the observed violation of Ellis-Jaffe sum rule \cite{ej}, and
giving a possible solution to the spin crisis problem \cite{emc}.

In this framework the quark distributions are parameterized in terms of
Fermi-Dirac statistical functions as
\bee
q_{a}(x) =  f(x)~\left[ \exp\left({ x - \tilde{x}(q_{a}) \over \bar{x}}\right)
+ 1\right]^{-1}~~~.
\label{eq:pparam}
\ene
Here $\tilde{x}(q_{a})$ plays the role of the {\it thermodynamical potential},
$\bar{x}$ of the temperature and $f(x)$ is the {\it level}-density in the $x$
variable. This function is ultimately related to the non perturbative
dynamics responsible for the binding of quarks and gluons inside the
hadrons, so it is theoretically undetermined. Analogously for the gluons (we
neglect their polarization) the Bose-Einstein relation has been assumed
\bee
G(x) = { 16 \over 3}f(x)~
\left[ \exp\left({ x - \tilde{x}(G) \over \bar{x}}\right)
- 1\right]^{-1}~~~,
\label{eq:gparam}
\ene
where now the factor $16/3$ is due to the colour degeneracy with respect to the
quarks case and to the sum over the two helicity states. Notice that the weight
function $f(x)$ has been assumed universal, being the same in (\ref{eq:pparam})
and (\ref{eq:gparam}). Moreover, the previous considerations allow also to
assume the relation
\bee
d(x) = \frac{u^{-}(x)}{1-F}~~~,
\label{eq:dappr}
\ene
and a dipole approximation for the $d$-quark polarization, namely
\bee
\Delta d(x) = - k~f(x)~\exp\left( {x - \tilde{x}(u^-)
\over \bar{x}}\right) \left[ \exp\left( {x - \tilde{x}(u^-)
\over \bar{x}}\right) + 1\right]^{-2}~~~.
\label{eq:deltad}
\ene
In terms of this parameterization it is possible to reproduce the NMC data
\cite{nmc} for $F^{p}_{2}(x)$ and $F^{n}_{2}(x)$ taken at $Q^2 = 4~GeV^2$.
These predictions are also compatible with the antiquark data obtained from
neutrino deep inelastic scattering \cite{antiq}, and the results known for the
gluon distribution \cite{g1}, \cite{g2}. In particular, from the fit procedure
it has been obtained $A=0.579$, $\alpha=-0.845$, $k=0.769$, $\bar{x}=0.132$,
$\tilde{x}(u^+)=0.524$, $\tilde{x}(u^-)=0.143$, $\tilde{x}(\bar{u}^+)=-0.216$,
$\tilde{x}(\bar{d}^+)=\tilde{x}(\bar{d}^-)=\tilde{x}(\bar{u}^-) =-0.141$, and
$\tilde{x}(G)=-0.012$ \cite{bour}, where for the function $f(x)$ the following
form was assumed
\bee
f(x) = A~x^{\alpha}~~~,
\label{eq:densityf}
\ene
to match the singular behaviour of parton distribution at low $x$. In Figures 1
and 2 we show the good agreement between the theoretical predictions of this
model and the experimental data for $F^{p}_{2}(x) - F^{n}_{2}(x)$ and
$F^{n}_{2}(x) / F^{p}_{2}(x)$, whereas, in Figures 3 and 4 we report the
predictions for the polarized structure functions $x~g_{1}^{p}(x)$ and
$x~g_{1}^{n}(x)$, which fits quite well with the experimental data \cite{emc},
\cite{e142}.

The analysis performed so far is at fixed $Q^2=4~GeV^2$. In order to
consider also the experimental data  at different $Q^2$ a scaling evolution
equation is needed. In the large $x$ and $Q^2$ regime, AP equations provide a
reliable description for scaling \cite{ap}. As already mentioned in the
Introduction,
the low $x$ region is characterized by an overdense parton medium, so one has
to expect nonlinear effects in the evolution equations due to the overlapping
of parton wave-functions. Thus, a quite natural conclusion is that a set of
{\it generalized} AP equations which would describe the evolution in the
moderately low $x$ region, should take into account quantum statistical
effects. A way to approach this problem is to start from the analogy showed in
\cite{coll} occurring between standard leading-log AP equations and Boltzmann
transport equations.

\section{Altarelli-Parisi evolution equations as a set of Boltzmann
equations}

As well-known, the logarithmic dependence on $Q^2$ of the parton distribution
momenta, predicted in the framework of perturbative $QCD$, has a simple and
beautiful interpretation in terms of evolution equations for parton
distribution functions \cite{ap}. At leading-log level, the AP equations can be
written in the following way
\bee
{d \over {dt}}p_{A}(x,t)= {\alpha_{s}(t) \over 2 \pi} \int_{x}^{1} {dy \over y}
\sum_{B} p_{B}(y,t) P_{AB}\left( {x \over y} \right)~~~,
\label{eq:ap}
\ene
where $t = \ln(Q^{2}/\mu^{2})$, $\mu$ is some renormalization
scale and $p_{A}(x,t) $  denote the parton distribution functions
($A,B=$quarks, antiquarks and gluons). By defining
\bee
\tau \equiv \frac{1}{2 \pi b} \ln \left[ \frac{\alpha_{s}(0)}{\alpha_{s}(t)}
\right]~~~,
\label{eq:tau}
\ene
with $b \equiv (33-2n_{f})/(12 \pi)$ ($n_{f}$ is the number of flavours),
Eq. (\ref{eq:ap}) becomes
\bee
{d \over {d\tau}}p_{A}(x,\tau)= \int_{x}^{1} {dy \over y}
\sum_{B} p_{B}(y,\tau) P_{AB}\left( {x \over y} \right)~~~.
\label{eq:apb}
\ene
Note that the dependence on $\tau$ of r.h.s. of (\ref{eq:apb}) comes only
through $p_{B}(y,\tau)$. In Eqs. (\ref{eq:ap}) and (\ref{eq:apb}),
$P_{AB}(x/y)$ stand for the
splitting functions, evaluated by using standard {\em equivalent parton
method}. They correspond to the {\it probability} for the elementary three-body
processes to occur in which a parton with momentum fraction $x$ is produced
by a parton with higher fraction $y=x/z$. Following the original
Altarelli-Parisi approach, the $1/(1-z)$ singularities of $P_{AA}$,
are removed by introducing the $(1-z)_{+}$ regularization
prescription\footnote{The integrals are defined in this prescription by
$\int_{0}^{1} [f(z)/(1-z)_{+}]~dz \equiv
\int_{0}^{1} [f(z)-f(1)]/(1-z)~dz$}, which explicitly implements the
cancellation occurring between the real and virtual soft gluons
emissions. In this way one gets
\begin{eqnarray}
P_{qq} &=& {4 \over 3} \left[{ 1 + z^2 \over (1 -z)_{+}}
+ { 3 \over 2} \delta(1-z)\right]~~~,\label{eq:PAB1}\\
P_{qg} &=& {1 \over 2} [z^2 + (1 -z)^2]~~~,\label{eq:PAB2}\\
P_{gq} &=& {4 \over 3} {1 + (1-z)^2 \over z}~~~,\label{eq:PAB3}\\
P_{gg} &=& 6 \left[ z(1-z) + {1-z \over z} + {z \over (1-z)_{+}} \right]
+ 2 \pi b~\delta(1-z)~~~.\label{eq:PAB4}
\end{eqnarray}
The microscopic picture beyond the scaling violation equations
(\ref{eq:apb}), however, has one main difficulty: the splitting
functions (\ref{eq:PAB1})-(\ref{eq:PAB4}) cannot be all interpreted, strictly
speaking, as probability densities, since they are not positive
definite (e.g. $\int_{0}^{1} P_{qq}(z)~dz=0$). This, in particular, is
the reason for not having explicitly, in the r.h.s. of (\ref{eq:apb}),
terms corresponding to {\it inverse} processes, in which a parton with
$x$ momentum fraction ends up in others with smaller momenta.

An alternative representation of (\ref{eq:apb}) has been developed in
\cite{coll}, where their microscopical interpretation is more clear.
According to this, instead of using direct $1/(1-z)_{+}$
regularization, it is shown how the virtual diagrams, responsible for
parton wave function renormalization, are equivalent to real diagrams
with negative sign. Using in fact Mueller {\it cut-vertices}
technique \cite{muell}, in addition to the ordinary real diagrams,
leading to positive contribution to parton distributions variation, a
negative term arises, corresponding to virtual gluon
emission diagrams with exactly
the same form for the unregularized parton splitting function as that
of the real ones.

By helicity conservation at the quark-gluon vertex, and assuming $n_{f}$
different flavours for quarks ($j=1,...,n_{f}$) with two helicity states
($\lambda=+,-$), the evolution
equations for polarized quark distribution functions can be cast in
the following form
\beqa
 { d \over { d \tau} } q_{j\lambda}(x, \tau) & = & \int_x^1 { dz \over z}
\left\{ \gamma_{qq}(z)~ q_{j\lambda} \left( { x \over z }, \tau \right)
+  { 1 \over 2 } \gamma_{qg}(z)~ G \left( { x \over z }, \tau \right)
\right\}  \nonumber \\
&- &  q_{j\lambda}(x, \tau) \int_0^1  dz~  \gamma_{qq}(z)~~~.
\label{eq:apquark}
\enqa
Note that, for simplicity, we have assumed $G_{+} = G_{-} = G/2$.
We will come back on this point in the following.
The equations for antiquarks are easily obtained by the previous one by
substituting $q_{j\lambda} \leftrightarrow \bar{q}_{j \lambda}$.
Similarly for the gluon unpolarized distribution $G(x,\tau)$ one has
\beqa
{ d \over { d \tau} } G (x, \tau) &=& \int_x^1 { dz \over z}
\left\{ \gamma_{gg}(z)~ G \left( { x \over z }, \tau \right)
+ \sum_{j=1}^{n_{f}} \sum_{\lambda=+,-} \gamma_{gq}(z)~
\left[ q_{j\lambda} \left( { x \over z }, \tau \right)
+  \bar{q}_{j\lambda} \left( { x \over z }, \tau \right) \right] \right\}
\nonumber \\
& - &  { 1 \over 2} G(x, \tau) \int_0^1  dz \left[ \gamma_{gg}(z)
+ 2 n_{f} \gamma_{qg}(z) \right]~~~.
\label{eq:apgluoni}
\enqa
In the previous equations the splitting functions $\gamma_{AB}$ are
defined by
\begin{eqnarray}
\gamma_{qq} &=& {4 \over 3} { 1 + z^2 \over 1 -z}~~~,\label{eq:gam1}\\
\gamma_{qg} &=& {1 \over 2} [z^2 + (1 -z)^2]~~~,\label{eq:gam2}\\
\gamma_{gq} &=& {4 \over 3} {1 + (1-z)^2 \over z}~~~,\label{eq:gam3}\\
\gamma_{gg} &=& 6 \left[ z(1-z) + {1-z \over z} + {z \over 1-z}
\right]~~~.\label{eq:gam4}
\end{eqnarray}
Note that all the divergences due to the singular behaviour of $\gamma _{AB}$
are explicitly cancelled once the terms with opposite sign,
occurring in (\ref{eq:apquark}) and (\ref{eq:apgluoni}), are taken into
account.

As already mentioned, AP equations in the form (\ref{eq:apquark})
(\ref{eq:apgluoni}) have a very clear and intuitive physical
interpretation. Let us consider, for example, the $Q^2$
variation of a quark distribution function with momentum fraction $x$,
$q_{j\lambda}(x)$: from (\ref{eq:apquark}) we see that this is due to
two terms, with opposite sign. The first one corresponds to the
production of quarks with momentum $x$ from partons (gluons or quarks)
with higher momentum  fraction $y \geq x$, so it contributes with a
positive sign. On the contrary, the second term accounts for the
depletion of $x$ fraction quarks due to their decay in a quark with
smaller momentum fraction $z x$ plus a gluon.
In particular, notice that the
same probability densities $\gamma_{AB}$ appear in both contributions,
showing the physical soundness of the picture. Similar is the interpretation
of (\ref{eq:apgluoni}).

As remarked in \cite{coll} this formulation also allows for an
intriguing formal interpretation of AP equations as a set of transport
Boltzmann equations if one regards the variable $\tau$ as the
analogous of {\it time} variable. From this point of view the
evolution in $Q^2$ of parton densities appear to be strictly
equivalent to the evolution in {\it time} of statistical distributions
corresponding to interacting particles, forming a system approaching
equilibrium. To better illustrate this analogy it is useful to briefly
review the Boltzmann transport equation formalism.

\section{Statistical effects on parton distribution scaling behaviour}

As well-known, the Boltzmann set of equations describes the evolution to
equilibrium states of systems composed by many particles of several species
($i$ specie-index $i = 1,..,n$)
mutually interacting \cite{degr}. Assuming for simplicity
particles homogeneously and isotropically
distributed, we can define the numerical distribution functions as
\bee
n_{i}(\epsilon, t) \equiv g_{i}(\epsilon)f_{i}(\epsilon,t)~~~,
\label{eq:ni}
\ene
with $\epsilon$ denoting the energy, $f_{i}(\epsilon,t)$ the statistical
functions (they recover the usual Bose/Einstein or Fermi/Dirac at the thermal
equilibrium), and $g_{i}(\epsilon)$ the level-densities (weights)
corresponding to $\epsilon$. These last quantities should be fixed from the
beginning,
by studying the hamiltonian of the system. From (\ref{eq:ni}) follows
the expression for the total number-density of i-particles
\bee
N_{i}(t) = \int {d^3\vec{p} \over (2 \pi)^3}~
g_{i}(\epsilon)f_{i}(\epsilon,t)~~~,
\label{eq:nit}
\ene
where $\vec{p}$ is the $3$-momentum, with $\vec{p}^2=\epsilon^2-m^2$.
By using Eq. (\ref{eq:ni}), the Boltzmann equations can be cast in
the following form
\bee
{\cal L}~ n_{i} = C_{i}[{\bf f},{\bf g}] = C^{+}_{i}[{\bf f },{\bf g}]-
C^{-}_{i}[{\bf f},{\bf g}]~~~~~~~~~~~~~~~~~i=1,....,n~~~~,
\label{eq:liouv}
\ene
where ${\bf f} \equiv (f_{1}, ..., f_{n})$, ${\bf g} \equiv (g_{1}, ...,
g_{n})$, ${\cal L}$ is the Liouville operator, and $C_{i}[{\bf f},{\bf g}
]$ is the so
called collisional integral for the i-th  particle specie. The latter
is given by a thermal average of all possible processes which
change the density of the i-th specie.
Notice that in Eq. (\ref{eq:liouv}) we have defined $C^{+}_{i}[{\bf f},{\bf g}
]$ and $C^{-}_{i}[{\bf f},{\bf g}]$  as the contributions corresponding to the
interaction processes which create or destroy the i-th particle specie
respectively.
For  simple three body processes $A \rightarrow B+C$, $B \rightarrow A+C$,
if we are interested in describing, for example,
the modification of $B$ population,
the corresponding terms in $C_{B}[{\bf f},{\bf g}]$ are the following
\beqa
C^{+}_{B}[{\bf f},{\bf g}] & - & C^{-}_{B }[{\bf f},{\bf g}]   =
\int \int
{d^3 \vec{p}_{A} \over 2\epsilon_{A}}~
{d^3 \vec{p}_{C} \over 2\epsilon_{C}}~
\Bigr\{ |{\cal M}(A \rightarrow B+C)|^2
{}~{\delta(\epsilon_{A}-\epsilon_{B}-\epsilon_{C}) \over (2 \pi)^2} \nonumber
\\
&\times &~ \delta^{3}(\vec{p}_{A}-\vec{p}_{B}-\vec{p}_{C})~
n_{A}(\epsilon_{A},t) ~g_{B}(\epsilon_{B})~\left[1 \pm f_{B}(\epsilon_{B},t)
\right]~ g_{C}(\epsilon_{C})~\left[1 \pm f_{C}(\epsilon_{C},t)\right] \Bigr\}
\nonumber \\
& - &\int \int {d^3 \vec{p}_{A} \over 2\epsilon_{A}}~
{d^3 \vec{p}_{C} \over 2 \epsilon_{C}}~
{}~\Bigr\{ |{\cal M}(B \rightarrow A+C)|^2
{}~{\delta(\epsilon_{B}-\epsilon_{A}-\epsilon_{C}) \over (2 \pi)^2}
\delta^3(\vec{p}_{B}-\vec{p}_{A}-\vec{p}_{C}) \nonumber \\
&\times &~
n_{B}(\epsilon_{B},t) ~g_{A}(\epsilon_{A})~\left[1 \pm f_{A}(\epsilon_{A},t)
\right]~ g_{C}(\epsilon_{C})~\left[1 \pm f_{C}(\epsilon_{C},t)\right] \Bigr\}
\label{eq:coll}
\enqa
where $|{\cal M}|^2$ are the squared
moduli of transition amplitudes and the sign in the final state factors is
positive/negative depending on the bosonic/fermionic nature of particles.
In the limit of very small $f_{i}$ one has $(1\pm f_{i}) \sim 1$ and
the collisional term for very dilute systems is recovered.

Coming back to the analogy between AP equations and Boltzmann equations
outlined in the previous section, it is physically reasonable to
imagine that the AP evolution equations have to be
modified for sufficiently low $x$. In this regime the nucleons are filled with
a large number of quark-antiquark pairs and gluons (the sea) and thus, to take
into account in the correct way the presence of this large number of partons,
the decay processes should be considered in presence of a
surrounding plasma of both Fermi and Bose particles. Corrections induced by
quantum statistical effects to the scaling behaviour dictated by
{\it standard} AP equations
are therefore generally present, and in particular we expect that:
\begin{itemize}
\item[a)] Pauli blocking will suppress the production of quarks and
antiquarks with fraction $x$ corresponding to $filled$ levels;
\item[b)] the gluon emission probability through bremsstr\"{a}hlung
processes, considered in the standard picture leading
to AP equations, will be enhanced by the contribution
of induced-emission in presence of a rather relevant
number of gluons in the sea.
\end{itemize}
These effects would favour the production of gluon-quark pairs with larger
values of $x$ for the quarks and a smaller one for the gluon. Moreover the
gluon conversion processes in $q-\bar{q}$ pairs are expected to be reduced.

As shown in (\ref{eq:liouv}), (\ref{eq:coll})
in non-equilibrium
statistical mechanics all these effects are simply included by multiplying
the amplitudes modulus squared of the relevant processes, appearing in the
collisional integral, by the factors $1-f$
or $1+f$ for each Fermi or Bose particle in the final state, with $f$
denoting the particle distribution functions without any level-density
factor.
In equilibrium conditions these $f$ reach the standard stationary
Fermi-Dirac or Bose-Einstein form, while in general they depend on time.
Thus, it is reasonable to expect that similar factors should be introduced in
the {\it generalized} AP equations. In other words, standard AP equations
correspond to a set of Boltzmann equations for
a dilute system of partons, where statistical
effects can be neglected: for higher parton densities, if we assume that this
analogy still holds\footnote{ It seems to us that the microscopical and
{\it fundamental} character of the interpretation of AP equations as transport
equations supports this assumption.}, it follows that these effects, which
are present in transport equations, should be present in scaling equations as
well.

In the same spirit of (\ref{eq:pparam}) and (\ref{eq:gparam})
\cite{bour}, we will parametrize the quark,
antiquark and gluon distributions as
\begin{eqnarray}
q_{j\lambda}(x, \tau) & = & g_{j\lambda}(x)~f_{j}^{\lambda}(x, \tau)~~~,
\label{eq:prod1}\\
\bar{q}_{j\lambda}(x, \tau) & = &
\bar{g}_{j\lambda}(x)~\bar{f}_{j}^{\lambda}(x, \tau)~~~,\label{eq:prod2}\\
G(x, \tau) & = & g_{G}(x)~f_{G}(x, \tau)~~~,
\label{eq:prod3}
\end{eqnarray}
where $g_{j\lambda}(x)$, $\bar{g}_{j\lambda}(x)$ and $g_{G}(x)$
are weight functions, whereas
$f_{j}^{\lambda}(x,\tau)$, $\bar{f}_{j}^{\lambda}(x,\tau)$
and $f_{G}(x,\tau)$ are purely statistical distributions, which
depending on $\tau$ cannot be assumed in principle to have equilibrium
form. The explicit form for $g$-functions, which contains
the divergency at
$x=0$, should be fitted from experimental data, as in \cite{bour}, or deduced
from theoretical expected behaviour, like, for example, Regge theory.
We stress that the factorized form (\ref{eq:prod1})-(\ref{eq:prod3}),
in particular the
hypothesis that the singular functions $g_{j\lambda}$, $\bar{g}_{j
\lambda}$ and
$g_{G}$ do not depend on $\tau$ is compatible with
predictions of both Regge theory and $QCD$ for the behaviour of parton
distributions at the end-point $x=0$. As it is well-known,
in this regime one has
\bee
p_{A}(x, Q^2) \sim \xi_{A}(Q^2) x^{- \alpha_{A}}~~~,
\label{eq:regge}
\ene
with $\alpha_{A}$ not depending on $Q^2$, at least for large $Q^2$
\cite{ynd}.

Within the factorized expression (\ref{eq:prod1})-(\ref{eq:prod3})
the final state factors
are written in the form $ 1 - f_{j}^{\lambda}$, $ 1 - \bar{f}_{j}^{\lambda}$
and $ 1+f_{G}$ for quarks, antiquarks and gluons respectively.

We are now able to introduce a set of generalized scaling equations for
quarks and gluons. Here we will consider for simplicity
the case in which the gluons are supposed
not to have a significant net polarization in the nucleons with respect to the
one carried by quarks. We will assume, therefore $ G_{+}(x, \tau)=
G_{-}(x, \tau) = G(x, \tau)/2$.
It should be pointed out that this approximation is consistent with the
results obtained in \cite{bour} and \cite{buc}, where it is argued that
Pauli principle plays the essential role to generate the polarization of the
quark sea. This approximation is instead less
satisfactory in the framework of the different interpretation of the violation
of Ellis-Jaffe sum rule based on the
axial-vector current anomaly \cite{anom}.
This latter case, in fact, would require a very large
gluon polarization, i.e. $\Delta G = G_{+} - G_{-} \sim 3 \div 4$.
Notice however that, as shown in \cite{bour}, gluons are expected to be more
numerous than quarks, due to their Bose nature, so in any case one has
$\Delta G /G << \Delta q/q$, which supports our approximation.

By helicity conservation at the quark-gluon vertex, it is easily seen that
generalized evolution equations for polarized quark distribution functions
get the following form \cite{Fermilab}
\beqa
 { d \over { d \tau} } q_{j\lambda}(x, \tau) & = & \int_x^1 { dz \over z}
\left\{ \gamma_{qq}(z)~ q_{j\lambda} \left( { x \over z }, \tau \right)
\left[ 1 -  f_{j}^{\lambda} (x, \tau)\right]
\left[ 1 +  f_{G} \left( x \left(
 { 1 \over z} -1 \right), \tau \right) \right] \right. \nonumber \\
& +& \left.  { 1 \over 2 } \gamma_{qg}(z)~ G \left( { x \over z }, \tau \right)
\left[ 1 -  f_{j}^{\lambda} (x, \tau)\right]
\left[ 1 -  \bar{f}_{j}^{-\lambda} \left( x \left(
{ 1 \over z} -1 \right), \tau \right) \right] \right\}  \nonumber \\
&- &  q_{j\lambda}(x, \tau) \int_0^1  dz ~\gamma_{qq}(z)~
\left[ 1 -  f_{j}^{\lambda} (xz, \tau)\right]
\left[ 1 +  f_{G} \left( x \left(
1 - z \right), \tau \right) \right]~~~.
\label{eq:quark}
\enqa
The equations for antiquarks are easily obtained by the previous one by
substituting $q_{j\lambda} \leftrightarrow \bar{q}_{j \lambda}$ and
$f_{j}^{\lambda} \leftrightarrow \bar{f}_{j}^{\lambda}$.
Similarly for the gluon distribution $G(x,\tau)$ one has
\beqa
{ d \over { d \tau} } G (x, \tau) &=& \int_x^1 { dz \over z}
\left\{ \gamma_{gg}(z)~ G \left( { x \over z }, \tau \right)
\left[ 1 +  f_{G} (x, \tau)\right]
\left[ 1 +  f_{G} \left( x \left(
{ 1 \over z} - 1 \right), \tau \right) \right] \right. \nonumber \\
& +& \sum_{j=1}^{n_{f}} \sum_{\lambda=+,-} \gamma_{gq}(z)~
\left[ 1 +  f_{G} (x, \tau) \right]
\left\{ q_{j\lambda} \left( { x \over z }, \tau \right)
\left[ 1 -  f_{j}^{\lambda} \left( x \left(
{ 1 \over z} -1 \right), \tau \right) \right] \right.
\nonumber\\
& + & \left. \left. \bar{q}_{j\lambda} \left( { x \over z }, \tau \right)
\left[ 1 -  \bar{f}_{j}^{\lambda} \left( x \left(
{ 1 \over z} -1 \right), \tau \right) \right]
\right\} \right\}
 \nonumber \\
& - &  { 1 \over 2} G(x, \tau) \int_0^1  dz \left\{ \gamma_{gg}(z)~
\left[ 1 +  f_{G} (xz, \tau) \right]
\left[ 1 +  f_{G} \left( x \left(
1 - z \right), \tau \right) \right]  \right. \nonumber \\
& +& \left.  \sum_{j=1}^{n_{f}} \sum_{\lambda=+,-}
\gamma_{qg}(z)~ \left\{
\left[ 1 -  f_{j}^{\lambda} (xz, \tau)\right]
\left[ 1 -  \bar{f}_{j}^{-\lambda} \left( x \left(
1 -  z \right), \tau \right) \right]  \right\} \right\}~~~.
\label{eq:gluoni}
\enqa
Note that also in this case, as in (\ref{eq:apquark}) and (\ref{eq:apgluoni})
the divergent contributions due to $\gamma_{AB}$ exactly cancel. These
generalized equations predict also a different, more complicated, evolution for
momenta. By taking Mellin transform of both sides of (\ref{eq:quark}) and
(\ref{eq:gluoni}), in fact, one sees that the standard scaling behaviour should
be corrected by terms quadratic and cubic in distribution functions, which are
not simply products of momenta of quarks and gluon densities.

Finally, as for the standard AP equations, the scaling behaviour for
unpolarized quark distributions can be obtained by simply considering the sum
$q_{j}(x,\tau) = q_{j+}(x, \tau) + q_{j-}(x,\tau)$ (the same holds for
antiquarks). Notice, however, that since the introduction of final state
statistical factors spoils the linearity of the equations, the evolution of
$q_{j}(x,\tau)$ will depend on both the polarized distribution functions and
not simply on their sum.

\section{Conclusions and remarks}

As suggested by some experimental results \cite{nmc}, \cite{emc}, the
Fermi/Bose nature of partons could sensibly affect the observable quantities in
deep inelastic scattering on nucleons. This idea, already successfully applied
in \cite{bour} and \cite{buc}, mainly motivates our paper, in which a set of
{\it generalized} scaling-law equations for parton distributions which take
into account quantum statistics effects are suggested.

It is quite natural to think that quantum statistics may modify the scaling
behaviour of parton distribution functions for rather small $x$ and high $Q^2$.
This regime is in fact characterized by a large number of partons, which
partially overlap their wave-functions, thus to correctly treat it one has to
think in terms of parton-plasma dynamics, and the expected modifications to the
standard AP evolution equations should have both dynamical (different
processes) and/or statistical nature (statistical correlation between the wave
functions).\\ The Gribov-Lipatov-Ryskin equations \cite{glr} represents a
successful attempt to describe this system. It focus the attention only on the
dynamical aspect of the problem, considering new interactions among partons
which introduce in tha scaling-law nonlinear terms of the parton distribution
functions. These processes, which become relevant with the increasing of the
density, differently from the one considered in the standard AP approach
involve two or more partons in the initial state (annihilation processes). In
this paper we have stressed a different but complementar aspect,
trying to introduce only the modifications to the
evolution equation which are of genuine quantum statistical origine.
Hence, a complete description of this region in the $x-Q^2$
plane should take into account both the results.

At low $x$, but still at high $Q^2$ (perturbative QCD regime), the
bremsstr\"{a}hlung processes, responsible at leading-log level for scaling
breaking, are likely supposed to occur in presence of such an overdense {\it
gas} of partons. In this case Pauli blocking and gluon stimulated emission play
a relevant role in parton distributions dynamics and thus in their scaling-law.
We have introduced both this statistical effects to obtain a {\it generalized}
evolution law, starting from the observation that an intriguing analogy holds
between AP equations, at leading-log, and a set of Boltzmann transport
equations for a dilute gas of partons \cite{coll}, where a simple function of
the scale variable $Q^2$ in AP equations plays the role of {\it time}
parameter. Extending this analogy also to the case of a dense system, which is
the case for the $x-Q^2$ region under study, one is naturally led to a new set
of evolution equations in which statistical factors $(1 -
f_{j\lambda}(x,\tau))$ or $( 1 + f_{G}(x,\tau))$ appear in the r.h.s. of the
evolution equation (collisional integral) to take into account the final state
of the emitted parton. This approach implicitely suggests to consider a parton
distribution $q_{j \lambda}(x,\tau)$ as the product of a pure weight factor $
g_{j \lambda} (x,\tau)$ connected to the level-density and independent of
$Q^2$, like for example suggested at low $x$ by the Regge theory \cite{ynd},
times the statistical distributions $f_{j\lambda}(x,\tau)$. According to this
analogy and by virtue of Boltzmann $H$ theorem, one would naturally expect that
the normalized parton distributions $f_{j\lambda}(x,\tau)$,
$\bar{f}_{j\lambda}(x,\tau)$ and $f_{G}(x,\tau)$ should approach stationary
Fermi and Bose  expressions as $Q^2$ increases. Remarkably, these conclusions
seem to agree with the phenomenological results obtained in \cite{bour} and
suggest that the {\it thermalization} process is rapid enough to essentially
reach the equilibrium conditions at $Q^2=4~GeV^2$. This question, together
with the new equation for momenta of distributions (no more linear and
simple like in AP equations) will be the subject for further publications.

Finally, we want to stress the difference of this approach with respect to the
way in which the occurrence of Pauli blocking effects are {\it perturbatively}
studied in the literature
\cite{ross}. The single, independent
parton picture,
which is at the basis of the improved parton model, is
only possible for quite large $x$, where the low density parton
fluid which fills the hadron, allows to neglect
the statistical correlations between partons due to the
overlapping of their wave functions. This is not the case when we move to the
low $x$ regime, and thus in
this region this treatment is not completely justified. Alternative
approaches, even if heuristic, as the one presented here, have
to be investigated, analyzing first of all their predictions.

\bigskip

{\large \bf Acknowledgements}

\bigskip

We thank Prof. Franco Buccella for encouraging this work and for his
valuable comments.

\newpage
{\large \bf Figure Captions}

\bigskip

\begin{itemize}
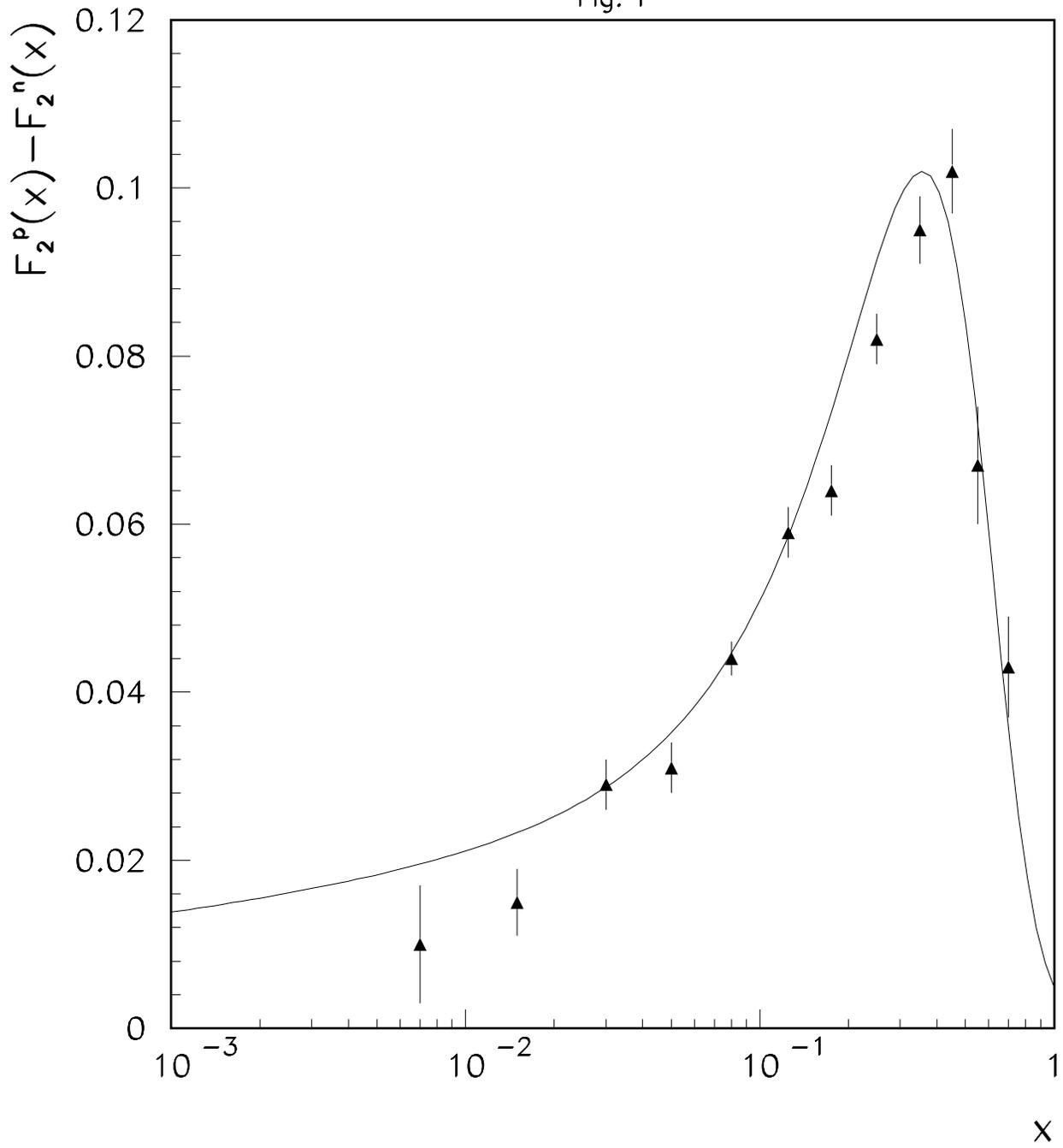
\item[Fig. 1.] The difference $F_{2}^{p}(x) - F_{2}^{n} (x)$ at $Q^2 = 4~GeV^2$
versus $x$. The experimental data are taken from \cite{nmc} and
the solid line represents the fit \cite{bour}.
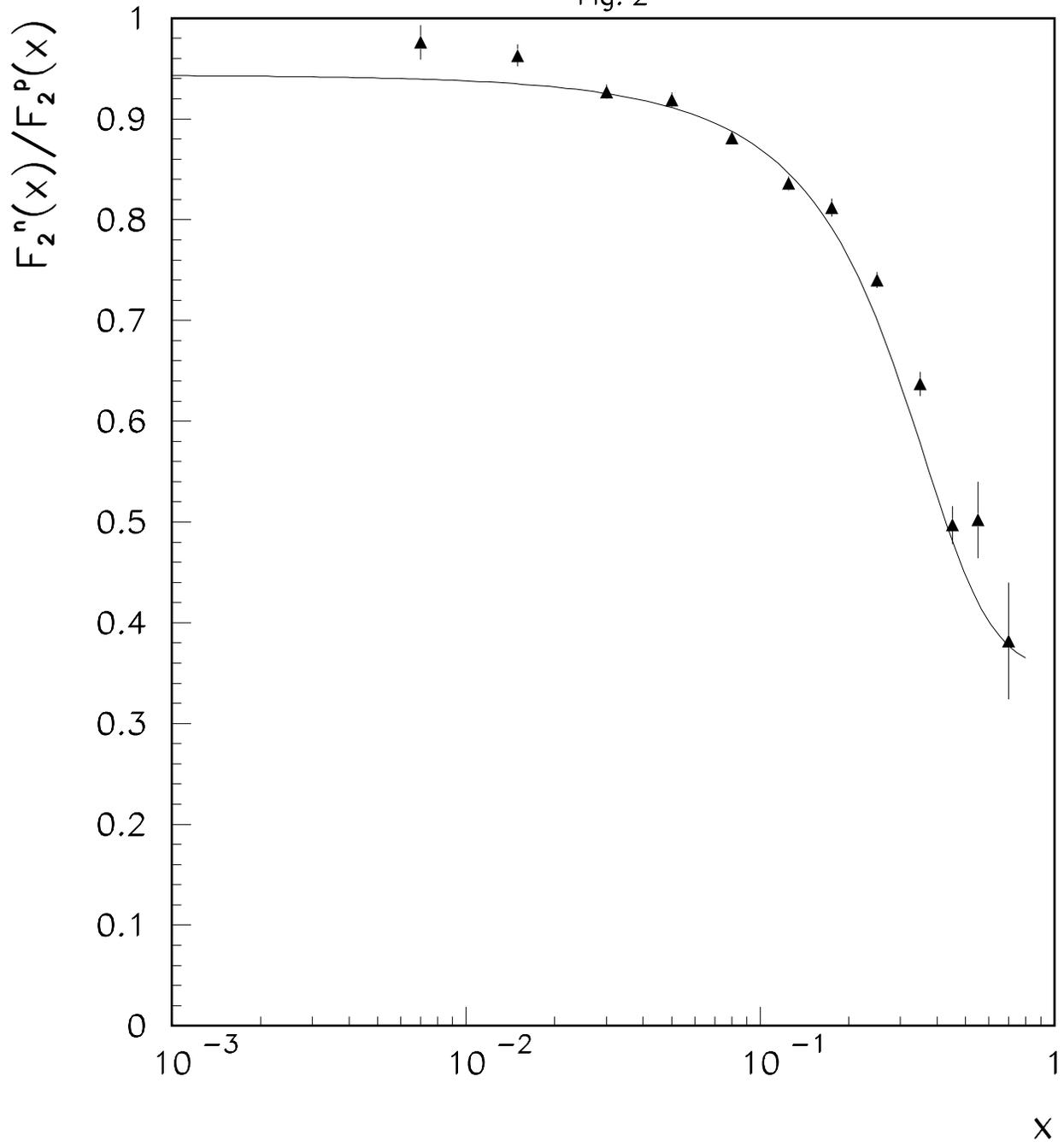
\item[Fig. 2.] The ratio $F_{2}^{n}(x)/F_{2}^{p} (x)$ at $Q^2 = 4~GeV^2$
versus $x$. The experimental data are taken from \cite{nmc} and
the solid line represents the fit \cite{bour}.
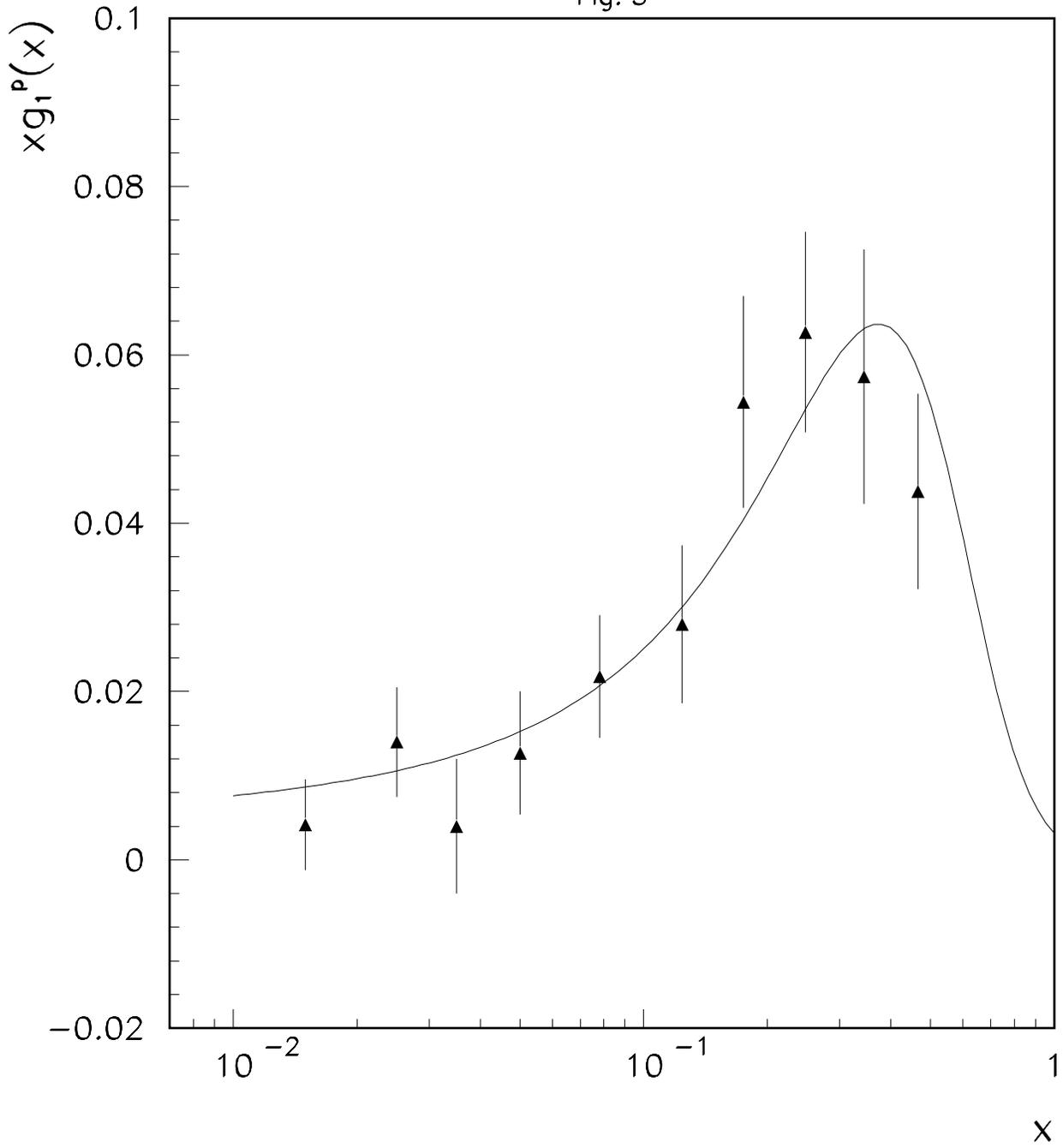
\item[Fig. 3.] $x~g_{1}^{p}(x)$ versus $x$. Data are from \cite{emc} and solid
line from \cite{bour}.
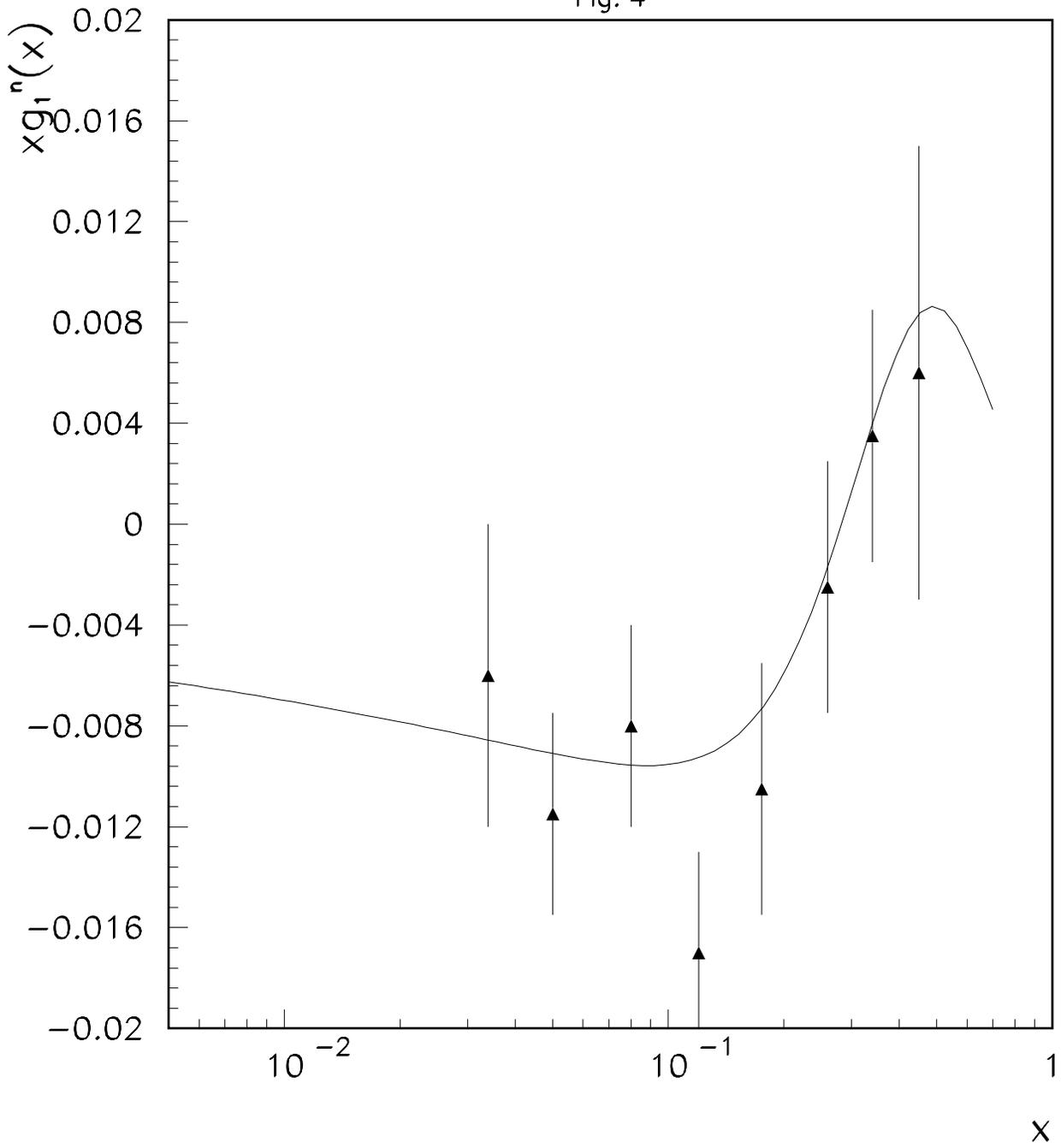
\item[Fig. 4.] $x~g_{1}^{n}(x)$ versus $x$. Data are from \cite{e142} and solid
line from \cite{bour}.
\end{itemize}

\end{document}